\documentclass[a4paper]{article}
\usepackage{a4wide}
\usepackage{graphicx}
\usepackage{amssymb}
\usepackage{amsmath}
\usepackage{amsfonts}
\usepackage{tikz}
\usepackage{pgfplots}
\usetikzlibrary{positioning}
\usepackage{booktabs}
\usepackage{clrscode3e}
\usepackage{enumerate}
\usepackage[shortlabels]{enumitem}
\usepackage{subfigure}
\usepackage[ruled,vlined]{algorithm2e}
\usepackage{multirow}
\usepackage{url}

\usepackage{natbib}
 \bibpunct[, ]{(}{)}{,}{a}{}{,}%

\DeclareMathOperator{\Direct}{\times} 
\DeclareMathOperator{\Kron}{\otimes}
 
\newcommand{\mat}[1]{\ensuremath{\mathbf{#1}}}
\newcommand{\tr}[1]{\ensuremath{#1}^\intercal}
\newcommand{\adj}[1]{\ensuremath{\mathbf{Adj(\mathnormal{#1})}}}
\newcommand{\R}{\mathbb{R}}
\newcommand{\B}{\mathbb{B}} 
\newcommand{\metricVar}{\textsc{var}}
\newcommand{\metricFrob}{\textsc{frob}}
\newtheorem{lemma}[section]{Lemma}
\newtheorem{definition}[section]{Definition}

\title{A Heuristic for Direct Product Graph Decomposition}

\author{Luca Calderoni, Luciano Margara, Moreno Marzolla\\
  Department of Computer Science and Engineering (DISI)\\
  University of Bologna\\
  {\small \texttt{luca.calderoni@unibo.it}, \texttt{luciano.margara@unibo.it}, \texttt{moreno.marzolla@unibo.it}}
}

\date{July 7, 2021}

\begin{document}

\maketitle

\begin{abstract}
  In this paper we describe a heuristic for decomposing a directed
  graph into factors according to the direct product (also known as
  Kronecker, cardinal or tensor product). Given a directed, unweighted
  graph~$G$ with adjacency matrix $\adj{G}$, our heuristic searches
  for a pair of graphs~$G_1$ and~$G_2$ such that $G = G_1 \Direct
  G_2$, where $G_1 \Direct G_2$ is the direct product of~$G_1$
  and~$G_2$.  For undirected, connected graphs it has been shown that
  graph decomposition is "at least as difficult" as graph isomorphism;
  therefore, polynomial-time algorithms for decomposing a general
  directed graph into factors are unlikely to exist. Although graph
  factorization is a problem that has been extensively investigated,
  the heuristic proposed in this paper represents -- to the best of
  our knowledge -- the first computational approach for general
  directed, unweighted graphs. We have implemented our algorithm using
  the MATLAB environment; we report on a set of experiments that show
  that the proposed heuristic solves reasonably-sized instances in a
  few seconds on general-purpose hardware.
\end{abstract}


\section{Introduction}\label{sec:intro}

Decomposition of complex structures into simpler ones is one of the
driving principles of mathematics and applied sciences. Everybody is
familiar with the idea of integer factorization, a topic that is
actively studied due to its number-theoretic as well as practical
implications, e.g., in cryptography. The concept of factorization can
be applied to other mathematical objects as well, such as graphs. Once
the concept of "graph product" is defined, one may naturally ask
whether a graph~$G$ can be decomposed into the product of two (or
more) smaller graphs.

Graph products are an active area of research because they are
involved in a number of computer science applications, such as load
balancing in distributed systems~\citep{load-balancing}, network
analysis~\citep{kronecker-graphs}, symbolic
computation~\citep{logspace-computations}, and quantum
computing~\citep{multiphoton}.  The most common types of graph
products that have been investigated in the literature are:
\emph{Cartesian product}, \emph{Direct product}, \emph{Strong product}
and~\emph{Lexicographic product}. Of these, the \emph{Direct product},
also known as~\emph{Kronecker} or~\emph{cardinal product}, is widely
used and will be the focus of this paper. We use the symbol $\Direct$
to denote the direct product, e.g., $G = G_1 \Direct G_2$.

It has recently been shown by~\cite{CALDERONI202172} that deciding
whether an undirected, unweighted, nonbipartite graph~$G$ is composite
according to the direct product, i.e., whether there exist nontrivial
graphs $G_1, G_2$ such that $G = G_1 \Direct G_2$, is at least "as
difficult as" deciding whether two graphs are isomorphic (a graph is
\emph{nontrivial} if it has more than one node). More formally, the
graph isomorphism problem is polynomial-time many-one reducible to the
graph compositeness testing problem (the complement of the graph
primality testing problem). A consequence of this result is that the
graph isomorphism problem for undirected, nonbipartite graphs is
polynomial-time Turing reducible to the primality testing problem. It
is therefore unlikely that there exists a polynomial-time algorithm
for graph factorization according to the direct product, unless graph
isomorphism is in~$P$.

The problem of graph factorization has been extensively
studied~\citep{hammack2011handbook} from the theoretical point of
view: mathematical properties of graph products are known, as well as
factorization algorithms for a few special cases. For example, it is
known that prime factorization of an undirected, connected and
non-bipartite graph with~$n$ nodes and~$m$ edges can be found in time
$O(mn^2)$~\citep{imrich1998}.  The result from~\cite{CALDERONI202172}
suggests that the lack of connectedness plays a major role in making
direct product primality testing and factorization harder.

Despite the large volume of theoretical work, the problem of graph
factorization has not received yet much attention from the
experimental research community. Indeed, to the best of our knowledge,
no implementation of graph factorization algorithms for general
directed graphs is available. The problem is exacerbated by the fact
that the existing algorithms only work on special kinds of graphs, and
it is not known whether they can be generalized to arbitrary directed
graphs, or whether different algorithms exist for general graphs.  In
this paper we begin to bridge the gap between theory and practice by
proposing a heuristic for direct-product factorization of general
graphs: given a directed, unweighted graph~$G$ with~$n$ nodes and two
positive integers $n_1, n_2$ such that $n = n_1 \times n_2$, our
algorithm finds two nontrivial graphs $G_1, G_2$ with~$n_1$ and~$n_2$
nodes, respectively, such that $G = G_1 \Direct G_2$, provided that
such graphs exist.

To the best of our knowledge, the algorithm described in this paper is
the first algorithm tackling the problem of factorization of general
unweighted graphs, i.e., graphs whose structure is not subject to any
constraint. Our algorithm implements a heuristic based on
gradient-descent local search. As such, it is not guaranteed that the
algorithm finds a solution even if one exists; however, we illustrate
a set of computational experiments that show that our algorithm does
find a valid solution quickly in many cases.

\section{Notation and Basic Definitions}\label{sec:preliminaries}

A directed graph $G = (V, E)$ is described as a finite set~$V$ of
nodes $V = \{v_1, \ldots, v_n\}$ and a finite set of edges $E
\subseteq \{(u, v)\ |\ u, v \in V\}$, where an edge $e \in E$ is an
ordered pair $e = (u, v)$, $u, v \in V$; an edge of the form $(v, v)$
is called \emph{self loop} or simply \emph{loop}. Given a graph~$G$,
$V(G)$ and~$E(G)$ are the set of nodes and edges of~$G$,
respectively. We denote by $G_1 \cup G_2$ the disjoint union of
graphs~$G_1$ and~$G_2$, i.e., the graph with node set $V(G_1) \cup
V(G_2)$ and edge set $E(G_1) \cup E(G_2)$; disjoint means that $V(G_1)
\cap V(G_2)=\emptyset$.

Four types of graph products have been investigated in the literature:
\emph{Cartesian product}, \emph{Direct product}, \emph{Strong product}
and \emph{Lexicographic product}. In all cases, the product of two
graphs $G_1, G_2$ is a new graph~$G$ whose set of nodes is the
Cartesian product of~$V(G_1)$ and~$V(G_2)$:

\begin{align*}
V(G) &= V(G_1) \times V(G_2) = \{ (u, v)\ |\ u \in V(G_1) \wedge v \in V(G_2) \}
\end{align*}

In this paper we are concerned with the \emph{Direct product}, also
known as Kronecker or cardinal product. The direct product of two
graphs $G_1, G_2$ is denoted as $G = G_1 \Direct G_2$, where $V(G) =
V(G_1) \times V(G_2)$ and
\begin{align*}
    E(G) &= \left\lbrace \left( (x, y), (x', y') \right)\ |\ (x, x') \in E(G_1) \wedge (y, y') \in E(G_2) \right\rbrace
\end{align*}

\begin{figure}[t]
  \centering
  \begin{tikzpicture}[scale=2,>=stealth,auto, minimum size=1.2cm]
    \node [draw , circle] (G11) at (0, 1) {$1$};
    \node [draw , circle] (G12) at (0, 2) {$2$};
    \path[] 
    (G11) [->] edge (G12)
    (G12) [->] edge [loop above] (G12);
    \node at (-0.5,1.5) {$G_1$};     
  
    \node [draw , circle] (G31a) at (1, 1) {$1,a$};
    \node [draw , circle] (G32a) at (1, 2) {$2,a$};
    \node [draw , circle] (G31b) at (2, 1) {$1,b$};
    \node [draw , circle] (G32b) at (2, 2) {$2,b$};
    \node [draw , circle] (G31c) at (3, 1) {$1,c$};
    \node [draw , circle] (G32c) at (3, 2)    {$2,c$};
    \path[] 
    (G31a) [->] edge (G32b)
    (G31b) [->] edge (G32a)
    (G31b) [->] edge (G32b)
    (G31c) [->] edge (G32b)
    (G32a) [->] edge [bend left=20] (G32b)
    (G32b) [->] edge [bend left=20] (G32a)
    (G32b) [->] edge [loop above] (G32b)
    (G32c) [->] edge (G32b);
    \node at (4, 1.5) {$G_1 \Direct G_2$};
    
    \node [draw , circle] (G2a) at (1, 0) {$a$};
    \node [draw , circle] (G2b) at (2, 0) {$b$};
    \node [draw , circle] (G2c) at (3, 0) {$c$};
    \path[] 
    (G2a) [->] edge [bend left=20] (G2b)
    (G2b) [->] edge [bend left=20] (G2a)
    (G2b) [->] edge [loop below] (G2b)
    (G2c) [->] edge (G2b);
    \node at (1.5,-0.5) {$G_2$};
\end{tikzpicture}
\caption{Direct product $G = G_1 \Direct G_2$}\label{fig:direct-product-example}
\end{figure}
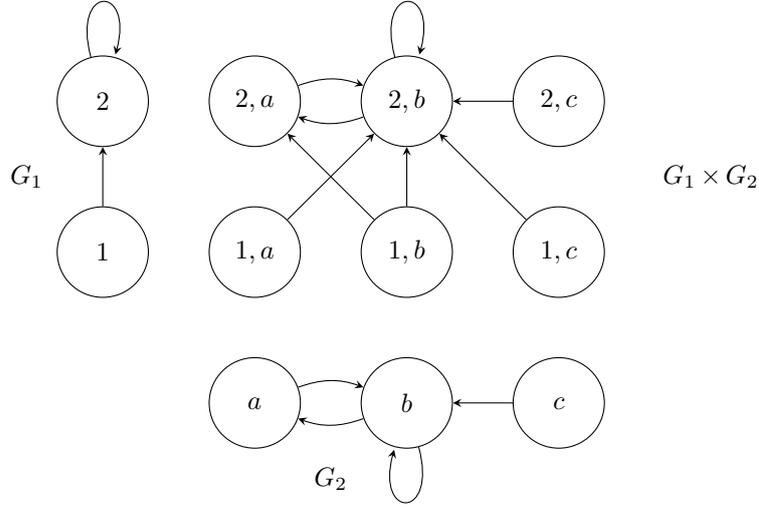

Figure~\ref{fig:direct-product-example} shows an example of direct
product of two graphs $G_1, G_2$.

The edge set of an unweighted graph~$G$ can be represented as an
adjacency matrix~$\adj{G}$. If~$G$ has~$n$ nodes, its adjacency matrix
$\mat{M} = \adj{G}$ is a $n \times n$ binary matrix, where $m_{ij} =
1$ if and only if $(v_i, v_j) \in E$. We denote the set of binary
matrices of size $n \times m$ as $B^{n \times n}$ the , $\B = \{0,
1\}$.

The algorithm described in the paper relies on the fact that the
direct product~$G_1 \Direct G_2$ can be expressed in terms of the
\emph{Kronecker product} of their adjacency matrices~\citep{Weischel,
  hammack2011handbook}. Given a $n \times m$ matrix~$\mat{B}$ and a $p
\times q$ matrix~$\mat{C}$, the Kronecker product $\mat{A} = \mat{B}
\Kron \mat{C}$ is a $np \times mq$ matrix that is composed of $n
\times m$ blocks of size $p \times q$, each block being the product of
elements of~$\mat{B}$ and the whole matrix~$\mat{C}$:

\begin{align}
  \begin{split}
    \mat{A} &= \mat{B} \Kron \mat{C} = \begin{pmatrix}
      b_{11} \mat{C} & b_{12} \mat{C} & \ldots & b_{1m} \mat{C} \\
      b_{21} \mat{C} & b_{22} \mat{C} & \ldots & b_{2m} \mat{C} \\
      \vdots & \vdots & \ddots & \vdots \\
      b_{n1} \mat{C} & b_{n2} \mat{C} & \ldots & b_{nm} \mat{C} \\
    \end{pmatrix}
  \end{split}\label{eq:kronecker}
\end{align}

Note that if $\mat{B}, \mat{C}$ are binary matrices, then $\mat{A} =
\mat{B} \Kron \mat{C}$ will be as well.  The relation between the
direct product of graphs and the Kronecker product of their adjacency
matrices is expressed by the following Lemma.

\begin{lemma}[\cite{CALDERONI202172}]\label{obs:kron-mat-kron-graph}
Given two directed, unweighted graphs~$G_1$ and~$G_2$, then
\begin{equation*}
\adj{G_1 \Direct G_2} = \tr{\mat{P}}\,(\adj{G_1} \Kron \adj{G_2})\,\mat{P}
\end{equation*}
where~$\mat{P}$ is a suitable permutation matrix.
\end{lemma}

We recall that a \emph{permutation matrix} $\mat{P} \in \B^{n \times
  n}$ is a square binary matrix with exactly a single~$1$ on each row
and column.  In other words, the adjacency matrix oi the direct
product of $G_1, G_2$ is equal to the Kronecker product of the
adjacency matrices of~$G_1$ and~$G_2$, up to a rearrangement
(relabeling) of the nodes of the resulting graph $G_1 \Direct G_2$.

For example, for the graphs in Figure~\ref{fig:direct-product-example}
we have:

\begin{align*}
\adj{G_1} &= \begin{pmatrix}
0 & 1 \\
0 & 1
\end{pmatrix} &
\adj{G_2} &= \begin{pmatrix}
0 & 1 & 0 \\
1 & 1 & 0 \\
0 & 1 & 0 
\end{pmatrix} &
\adj{G_1 \Direct G_2} &= \left(\begin{array}{ccc|ccc}
0 & 0 & 0 & 0 & 1 & 0 \\
0 & 0 & 0 & 1 & 1 & 0 \\ 
0 & 0 & 0 & 0 & 1 & 0 \\ \hline
0 & 0 & 0 & 0 & 1 & 0 \\ 
0 & 0 & 0 & 1 & 1 & 0 \\
0 & 0 & 0 & 0 & 1 & 0
\end{array}\right)
\end{align*}

In this case the permutation matrix is the identity matrix, since we
are assuming the mapping $(1,a) \rightarrow 1, (1, b) \rightarrow 2,
(1,c) \rightarrow 3, \ldots$ of nodes into indices. A different
mapping would require relabeling the nodes of $G_1 \Direct G_2$.

\begin{table}[t]
  \centering%
  \begin{tabular}{lp{.75\textwidth}}
    \toprule
    \textbf{Symbol} & \textbf{Meaning} \\
    \midrule
    $\B$            & The set $\{0, 1\}$ \\
    $\adj{G}$       & The adjacency matrix of graph~$G$ \\
    $\mat{P}$       & Permutation matrix \\
    $\Direct$       & The direct graph product operator \\
    $\Kron$         & The Kronecker matrix product operator \\
    $\phi(\mat{A}\, \mat{B}, \mat{C})$ & Metric that shows "how well" $\mat{A}$ can be
                                         written as the Kronecker product $\mat{B} \Kron \mat{C}$ (lower is better)\\
  \bottomrule
  \end{tabular}
  \caption{Summary of notation.}\label{tab:notation}
\end{table}

A nontrivial graph~$G$ is a graph with more than one node ($|V(G)| >
1$). We say that a graph~$G$ is \emph{prime} according to a given
graph product $\odot$ if~$G$ is nontrivial and $G = G_1 \odot G_2$
implies that either~$G_1$ or~$G_2$ are trivial, i.e., one of them has
exactly one node.

Finally, we introduce the concept of \emph{block matrix} that will be
used in the following to discuss several subroutines of our heuristic.

\begin{definition}\label{def:blockMatrix}
Let us consider a binary matrix~$\mat{A}$ made of $\mathit{dimB}
\times \mathit{dimB}$ submatrices (or blocks) $\mat{C}_{ij}$ of size
$\mathit{dimC} \times \mathit{dimC}$ each.  Let~$\mu$ be the the
average number of 1s on each block, and let~$s_{ij}$ be the number of
1s in $\mat{C}_{ij}$; we say that~$\mat{A}$ is a \emph{block matrix}
if and only if for each block~$\mat{C}_{ij}$:
\[
    s_{ij}=
    \begin{cases}
        x, & \mbox{if}\ x > \mu  \\
        0, & \mbox{otherwise}
    \end{cases}
\]
\end{definition}

In other words, a \emph{block matrix} is made of blocks such that each
block is either entirely zero, or has a number of 1s that is strictly
above the average number of 1s over all blocks.

Table~\ref{tab:notation} summarizes the notation used in this paper.

\section{An Heuristic for Direct Product factorization}\label{sec:algorithm}

In this section we describe a heuristic for decomposing a directed,
unweighted graph~$G$ into the direct product of two nontrivial graphs
$G_1, G_2$, if such graphs exist. We assume that the size (number of
nodes) of~$G_1$ and~$G_2$ is known; if this is not the case, then by
the definition of direct product it must hold that $n = n_1 \times
n_2$ where $n, n_1, n_2$ are the number of nodes of $G, G_1, G_2$
respectively; therefore, both~$n_1$ and~$n_2$ must be nontrivial
divisors of~$n$. Since the number of divisors of~$n$ is bounded from
above by~$n$, we can brute-force all combinations of $n_1, n_2$, whose
number grows polynomially w.r.t.~$n$.

The basic idea is to take advantage of
Lemma~\ref{obs:kron-mat-kron-graph} to find a suitable permutation
matrix~$\mat{P}$ so that the (permuted) adjacency matrix of~$G$ can be
written as the Kronecker product of two smaller matrices~$\mat{B}$
and~$\mat{C}$; these matrices can then be interpreted as the adjacency
matrices of two graphs $G_1, G_2$ with the property that $G = G_1
\Direct G_2$.  The problem is equivalent to finding a graph~$G'$
isomorphic to~$G$ such that $\adj{G'} = \mat{B} \Kron \mat{C}$; this
is not surprising, given the relationship between graph isomorphism
and compositedness testing proven in~\cite{CALDERONI202172}.

One more ingredient is needed to complete the heuristic, i.e., a way
to decide whether a binary square matrix can be written as the
Kronecker product of two smaller matrices. This is an instance of the
more general \emph{nearest Kronecker product} (NKP)
problem~\citep{VanLoan1993,ubiquitous-kronecker-product}: given
$\mat{A} \in \R^{m \times n}$ with $m = m_1 m_2$, $n = n_1 n_2$, find
$\mat{B} \in \R^{m_1 \times n_1}$ and $\mat{C} \in \R^{m_2 \times
  n_2}$ such that

\begin{equation}\label{eq:nkp}
\phi(\mat{A}, \mat{B}, \mat{C}) = \| \mat{A} - \mat{B} \Kron \mat{C} \|
\end{equation}

\noindent is minimized, according to some norm $\| \cdot \|$. If
$\mat{A}$ is the Kronecker product of~$\mat{B}$ and~$\mat{C}$, then
the minimum norm would be zero ($\phi(\mat{A}, \mat{B}, \mat{C}) =
0$). The NKP problem~\eqref{eq:nkp} can be solved by computing a
Singular Value Decomposition (SVD) of a suitably reshaped and permuted
version of matrix~$\mat{A}$ (see~\cite{VanLoan1993} for details).

To recap: to decompose a directed, unweighted graph~$G$ with adjacency
matrix $\mat{M} = \adj{G}$, we need to find a permutation
of~$\mat{M}$, say $\tr{\mat{P}} \mat{M} \mat{P}$, and two square
matrices $\mat{M}_1 = \adj{G_1}$ and $\mat{M}_2 = \adj{G_2}$ of given
sizes such that:

\begin{align*}
0 &= \phi(\tr{\mat{P}} \mat{M} \mat{P}, \mat{M}_1, \mat{M}_2) \\
  &= \| \tr{\mat{P}} \mat{M} \mat{P} - \mat{M}_1 \Kron \mat{M}_2 \|
\end{align*}

With a slight abuse of notation, in the following we write
$\phi(\mat{A})$ to denote the minimum value of $\phi(\mat{A}, \mat{B},
\mat{C})$ that can be achieved by suitably choosing $\mat{B},
\mat{C}$; in other words, $\phi(\mat{A})$ shows how well
matrix~$\mat{A}$ can be expressed as the Kronecker product of smaller
matrices~$\mat{B}, \mat{C}$ of given sizes.  $\phi(\cdot)$ is a
\emph{lower is better} metric: if $\phi(\mat{A}) = 0$, the
matrix~$\mat{A}$ is in Kronecker form.

Note that we must impose the additional constraint that~$\mat{M}_1$
and~$\mat{M}_2$ must be binary matrices, which is not guaranteed by
but the algorithm for Kronecker factorization described
in~\cite{VanLoan1993}. We will show below how we modified the NKP
algorithm to cope with this requirements.

\begin{algorithm}[ht]
\SetAlgoLined
\KwIn{$\mat{A} \in \B^{n \times n}$, $n = n_1 \times n_2$, $K$}
\KwOut{$\mat{P} \in \R^{n \times n}$, $\mat{B} \in \B^{n_1 \times n_1}$, $\mat{C} \in \B^{n_2 \times n_2}$ or "no solution found"}
$\mat{P}$ = Identity matrix of size $n \times n$\;
\Repeat{$(r_j < \epsilon) \vee (\textrm{number of iterations } \geq \textrm{threshold}$)}{
    \For{$i=1, \ldots, K$}{
        Let $\mat{P}_i$ be obtained from $\mat{P}$ by exchanging a random pair of rows/columns\;
        Find $\mat{B}_i \in \B^{n_1 \times n_1}, \mat{C}_i \in \B^{n_2 \times n_2}$ s.t. $r_i = \phi(\tr{\mat{P}_i} \mat{A} \mat{P}_i, \mat{B}_i, \mat{C}_i)$ is minimized\;
    }
    Let $j = \arg\min_i \{ r_i \}$\;
    $\mat{P} = \mat{P}_j$\;
}
\eIf{$r_j < \epsilon$}{
    \KwRet $\mat{P}, \mat{B}_j, \mat{C}_j$\;
}{
    \KwRet "no solution found"\;
}
\caption{Direct product decomposition}\label{alg:dir-prod-decomposition}
\end{algorithm}

Algorithm~\ref{alg:dir-prod-decomposition} shows a very high-level
overview of the proposed heuristic. The algorithm uses the
gradient-descent technique to find the permutation~$\mat{P}$ such that
the matrix $\tr{\mat{P}}\mat{A}\mat{P}$ can be decomposed according to
the Kronecker product.

Matrix~$\mat{P}$ is built incrementally, starting from the identity
permutation, by exchanges of rows and columns. At each step, the
algorithm tries to decrease the value of $\phi(\tr{\mat{P}} \mat{A}
\mat{P})$ so that either the value becomes zero (in which case we have
found a decomposition of the input graph), or no optimal permutation
is found within the allotted number of steps.

To limit the search space, at each step the algorithm generates~$K$
random permutations of the current matrix~$\mat{P}$; for each
permutation $\mat{P}_i$, the algorithm solves the NKP problem by
identifying two binary matrices $\mat{B}_i, \mat{C}_i$ such that

\begin{align*}
r_i &= \phi(\tr{\mat{P}_i} \mat{A} \mat{P}_i, \mat{B}_i, \mat{C}_i) = \| \tr{\mat{P}_i} \mat{A} \mat{P} - \mat{B}_i \Kron \mat{C}_i \|
\end{align*}

\noindent is minimized. If the minimum value~$r_j$ is not zero, the
process is repeated starting from the "best" permutation
matrix~$\mat{P}_j$ found so far.  If no solution is found after some
maximum number of iterations, the procedure assumes that
matrix~$\mat{A}$ can not be expressed as the Kronecker product of two
matrices of size $n_1 \times n_1$ and $n_2 \times n_2$.

Despite its appealing simplicity, a direct implementation of
Algorithm~\ref{alg:dir-prod-decomposition} is not effective in solving
the graph decomposition problem. First of all, gradient-descent
procedures may become stuck in a local minimum, with the result that
they fail to find a global optimum (in our case, the global optimum is
a permutation~$\mat{P}_j$ for which $r_j = 0$, provided that such a
permutation exists).  Another issue, already stated above, is that we
need to solve a modified version of the NKP problem, in which the
factors~$\mat{B}$ and~$\mat{C}$ must be binary matrices.

\paragraph{Algorithm Details}
We now provide the detailed description of our graph decomposition
heuristic, where both these issues will be addressed. The complete
pseudocode of the heuristic is provided in the Appendix, while a
MATLAB implementation along with the source code can be downloaded
from~\url{https://github.com/calderonil/kron}.

As already shown in~\eqref{eq:kronecker}, the Kronecker product
$\mat{A} = \mat{B} \Kron \mat{C}$ of two binary matrices~$\mat{B}$
and~$\mat{C}$ has a block structure: if $b_{ij} = 0$ then the
corresponding block $b_{ij} \mat{C}$ will contain only zeros.  The
algorithm consists of a main procedure --- \emph{alternateLocalSearch}
--- and by three subroutines. The main procedure, detailed in
Algorithm~\ref{alg:alternateLocalSearch}, swaps rows/columns and
evaluates if the swaps lead to an improvement (reduction) in the value
of function $\phi()$~\eqref{eq:nkp}.  We employ two different
metrics~\metricVar~and~\metricFrob~to evaluate~$\phi()$.

Let~$s_{ij}$ be the number of 1s in the block $\mat{C}_{ij}$,
\metricVar~computes the variance among the number of 1s of each block,
namely $\sigma^2 = Var\{s_{ij}\ \|\ 1 \leq i, j \leq
\mathit{dimB}\}$. When the variance increases, the matrix~$\mat{A}$ is
approaching the \emph{block matrix} form (refer to
Definition~\ref{def:blockMatrix}). Conversely, the
metric~\metricFrob~measures how much the nonempty blocks of~$\mat{A}$
fit the Kronecker form, i.e., whether or not their content is the
same; this metric is based on that used in~\cite{VanLoan1993} for
solving eh MKP problem: the rows of the same block of size
$\mathit{dimC} \times \mathit{dimC}$ of~$\mat{A}$ are concatenated,
and become a row of a new matrix~$\mat{F}$. $\mat{F}$ is then
multiplied with $\tr{\mat{F}}$ element by
element. Metric~\metricFrob~is the squared sum of the elements of this
product.

The local search procedure starts with metric~\metricVar, and switches
back and forth to~\metricFrob~when no significant improvement in the
value of~$\phi()$ is observed for a given number of iterations.  The
use of two different metrics is motivated by the empirical observation
that it speeds up convergence towards the solution, and helps to
escape local minima. When we find a swap that improves the value of
function~$\phi()$, we apply the swap to the current permutation matrix
$\mat{P}$ and proceed with the next iteration. The main loop
terminates when the number of iterations exceeds a user-defined
threshold, or when a decomposition is found.

\begin{figure}[t]
\centering\hspace*{-1cm}\begin{tikzpicture}
\node at (-5.5,3) { \includegraphics[scale=0.5]{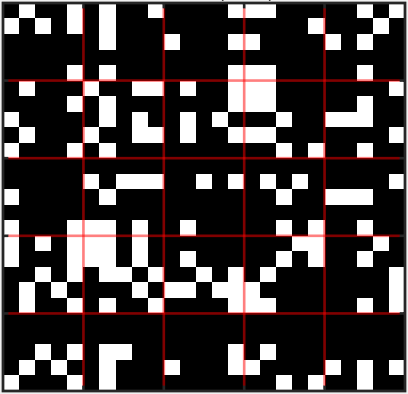} };%
\node at (0,3) { \includegraphics[scale=0.5]{img/kg1.png} };%
\node at (5.5,3) { \includegraphics[scale=0.5]{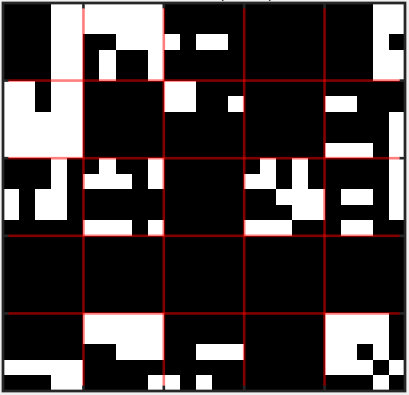} };%
\draw [thick, draw=black, fill=yellow, opacity=0.3] (-8.25,5.6) -- (-2.75,5.6) -- (-2.75,5.35) -- (-8.25,5.35) -- (-8.25,5.6);
\draw [thick, draw=black, fill=green, opacity=0.3] (-8.25,4.75) -- (-2.75,4.75) -- (-2.75,4.5) -- (-8.25,4.5) -- (-8.25,4.75);
\draw [thick, draw=black, fill=green, opacity=0.3] (-8.25,4.35) -- (-2.75,4.35) -- (-2.75,4.1) -- (-8.25,4.1) -- (-8.25,4.35);
\draw [thick, draw=black, fill=green, opacity=0.3] (-8.25,2.1) -- (-2.75,2.1) -- (-2.75,1.85) -- (-8.25,1.85) -- (-8.25,2.1);
\draw [thick, draw=black, fill=green, opacity=0.3] (-8.25,1.7) -- (-2.75,1.7) -- (-2.75,1.45) -- (-8.25,1.45) -- (-8.25,1.7);
\draw [thick, draw=black, fill=yellow, opacity=0.3] (-2.7,0.3) -- (-2.7,5.7) -- (-2.45,5.7) -- (-2.45,0.3) -- (-2.7,0.3);
\draw [thick, draw=black, fill=green, opacity=0.3] (-1.85,0.3) -- (-1.85,5.7) -- (-1.6,5.7) -- (-1.6,0.3) -- (-1.85,0.3);
\draw [thick, draw=black, fill=green, opacity=0.3] (-1.4,0.3) -- (-1.4,5.7) -- (-1.15,5.7) -- (-1.15,0.3) -- (-1.4,0.3);
\draw [thick, draw=black, fill=green, opacity=0.3] (0.95,0.3) -- (0.95,5.7) -- (1.2,5.7) -- (1.2,0.3) -- (0.95,0.3);
\draw [thick, draw=black, fill=green, opacity=0.3] (1.35,0.3) -- (1.35,5.7) -- (1.6,5.7) -- (1.6,0.3) -- (1.35,0.3);
\node at (-5.5,0) {$(a)$};
\node at (0,0) {$(b)$};
\node at (5.5,0) {$(c)$};
\end{tikzpicture}
\caption{kronGrouping procedure\label{fig:kronGrouping}. (a) and~(b):
  The current pivot (yellow) is compared with each other row and
  column of the matrix. (c) The $\mathit{dimC} - 1$ rows and columns
  that are most similar to the pivot (green) are selected as pivot
  neighbours in the current permutation.}
\end{figure}

Before swapping rows and columns, matrix $\mat{A}$ is processed by the
\emph{kronGrouping} procedure (Algorithm~\ref{alg:kronGrouping}). This
procedure tries to permute~$\mat{A}$ in such a way that it becomes
similar to a block matrix (see Figure~\ref{fig:kronGrouping}). The
intuition is that~$\mat{A}$ should be permuted in such a way that it
consists of blocks that contain as many 1s as possible, and others
that are all zero. \emph{kronGrouping} tries to exchange rows/columns
in such a way that nonempty blocks (i.e., blocks of $\mat{A}$ that
contain 1s) either lose or acquire 1s. To this end,
\emph{kronGrouping} evaluates the similarity between rows/columns;
given two vectors $\mat{v} = (v_1, \ldots, v_n)$ and $\mat{w} = (w_i,
\ldots, w_n)$, the similarity of $\mat{v}$ and $\mat{w}$ is a value
that is proportional to the number of elements for which $v_i = w_i$.
This information is used to produce a permutation that groups similar
rows/columns in the same submatrix.

\begin{figure}[t]
\centering\hspace*{-1cm}\begin{tikzpicture}
\node at (-5.5,3) { \includegraphics[scale=0.5]{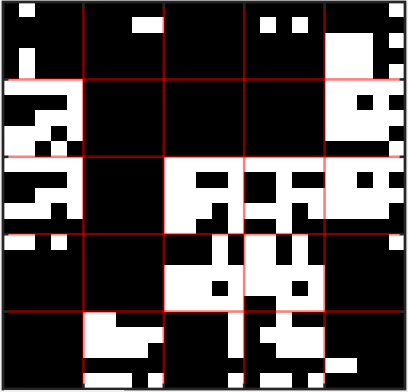} };%
\node at (0,3) { \includegraphics[scale=0.5]{img/out1.png} };%
\node at (5.5,3) { \includegraphics[scale=0.5]{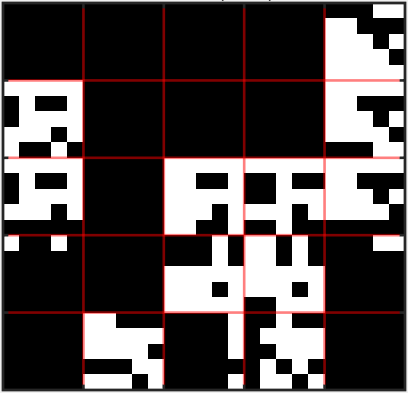} };%
\draw [thick, draw=black, fill=white, opacity=0.3] (-3.9,5.55) -- (-2.85,5.55) -- (-2.85,2.5) -- (-3.9,2.5) -- (-3.9,5.55);
\draw [thick, draw=black, fill=white, opacity=0.3] (-8.15,4.55) -- (-7.1,4.55) -- (-7.1,2.5) -- (-8.15,2.5) -- (-8.15,4.55);
\draw [thick, draw=black, fill=white, opacity=0.3] (-4.95,3.5) -- (-3.9,3.5) -- (-3.9,0.45) -- (-4.95,0.45) -- (-4.95,3.5);
\draw [thick, draw=black, fill=white, opacity=0.3] (-6.05,3.5) -- (-5,3.5) -- (-5,1.5) -- (-6.05,1.5) -- (-6.05,3.5);
\draw [thick, draw=black, fill=white, opacity=0.3] (-7.1,1.5) -- (-6.05,1.5) -- (-6.05,0.45) -- (-7.1,0.45) -- (-7.1,1.5);
\draw [thick, draw=black, fill=green, opacity=0.3] (-2.75,5.4) -- (2.75,5.4) -- (2.75,5.15) -- (-2.75,5.15) -- (-2.75,5.4);
\draw [thick, draw=black, fill=green, opacity=0.3] (-2.75,0.9) -- (2.75,0.9) -- (2.75,0.65) -- (-2.75,0.65) -- (-2.75,0.9);
\draw [thick, draw=black, fill=green, opacity=0.3] (-2.45,0.3) -- (-2.45,5.7) -- (-2.2,5.7) -- (-2.2,0.3) -- (-2.45,0.3);
\draw [thick, draw=black, fill=green, opacity=0.3] (2.2,0.3) -- (2.2,5.7) -- (2.45,5.7) -- (2.45,0.3) -- (2.2,0.3);
\node at (-5.5,0) {$(a)$};
\node at (0,0) {$(b)$};
\node at (5.5,0) {$(c)$};
\node at (-5.5,-3) { \includegraphics[scale=0.5]{img/out2.png} };%
\node at (0,-3) { \includegraphics[scale=0.5]{img/out2.png} };%
\node at (5.5,-3) { \includegraphics[scale=0.5]{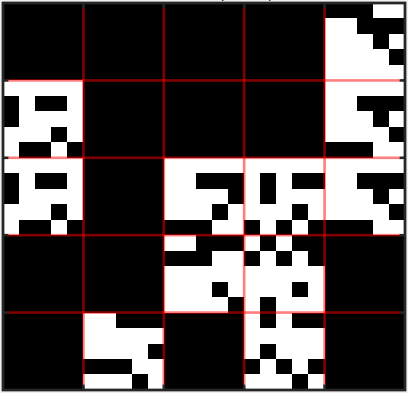} };%
\draw [thick, draw=black, fill=white, opacity=0.3] (-3.9,-0.45) -- (-2.85,-0.45) -- (-2.85,-3.5) -- (-3.9,-3.5) -- (-3.9,-0.45);
\draw [thick, draw=black, fill=white, opacity=0.3] (-8.15,-1.45) -- (-7.1,-1.45) -- (-7.1,-3.5) -- (-8.15,-3.5) -- (-8.15,-1.45);
\draw [thick, draw=black, fill=white, opacity=0.3] (-4.95,-2.5) -- (-3.9,-2.5) -- (-3.9,-5.55) -- (-4.95,-5.55) -- (-4.95,-2.5);
\draw [thick, draw=black, fill=white, opacity=0.3] (-6.05,-2.5) -- (-5,-2.5) -- (-5,-4.5) -- (-6.05,-4.5) -- (-6.05,-2.5);
\draw [thick, draw=black, fill=white, opacity=0.3] (-7.1,-4.5) -- (-6.05,-4.5) -- (-6.05,-5.55) -- (-7.1,-5.55) -- (-7.1,-4.5);
\draw [thick, draw=black, fill=green, opacity=0.3] (-2.75,-3.3) -- (2.75,-3.3) -- (2.75,-3.75) -- (-2.75,-3.75) -- (-2.75,-3.3);
\draw [thick, draw=black, fill=green, opacity=0.3] (0.3,-5.7) -- (0.3,-0.3) -- (0.75,-0.3) -- (0.75,-5.7) -- (0.3,-5.7);
\node at (-5.5,-6) {$(d)$};
\node at (0,-6) {$(e)$};
\node at (5.5,-6) {$(f)$};
\end{tikzpicture}
\caption{\emph{outsiders} procedure.\label{fig:outsiders}. The
  procedure detects those blocks that should gain or lose 1s. In (a)
  blocks that are going to acquire 1s are highlighted. In (b) we show
  the best swap is among rows/columns~$2$ and~$24$. The main loop will
  use this swap as first, resulting in an improvement with respect to
  the current metric. As such, the swap is effectively performed
  (c). The function \emph{outsiders} is called again at the next
  iteration and it selects $(15,16)$ as best swap (e). The swap is
  then applied; the matrix is now divided in blocks (f).}
\end{figure}

At each iteration of the main loop, \emph{alternateLocalSearch}
performs several operations before starting to test each possible
swap. First of all, the procedure checks whether~$\mat{A}$ is a block
matrix. If not, subroutine \emph{outsiders} is applied to~$\mat{A}$
(Algorithm \ref{alg:outsiders}). While \emph{kronGrouping} tries to
rearrange the whole matrix at a glance, \emph{outsiders} follows a
fine tuning perspective and derives an ordered list of swaps that seem
to be more convenient. This procedure allows to detect those rows and
columns that are evidently misplaced with respect to the block
structure we desire to reach.

To derive the swaps list, \emph{outsiders} verifies which block
of~$\mat{A}$ should preferably be filled (i.e. it has a number of 1s
above average) and which block should preferably be emptied. The
density of each block to be filled is considered as well: the best
swap is supposed to be the one that moves 1s from a block to be
emptied replacing zeros in a block to be filled. However, moving
elements in blocks that already have a high density would produce
overfilled blocks that are unlikely to appear in a feasible
factorization.

When the procedure terminates the \emph{bestSwaps} list is prepended
to the randomized list of each legal swap that is used in the main
loop as fallback. If the list of swaps is empty, there are no elements
set to one in each of the blocks that are likely to be emptied. Thus,
according to Definition~\ref{def:blockMatrix}, the matrix is now a
block matrix. The basic principles of \emph{outsiders} procedure are
outlined in Figure~\ref{fig:outsiders}.

\begin{figure}[t!]
\centering\hspace*{-1cm}\begin{tikzpicture}
\node[inner sep=0pt] at (-5.5,3) { \includegraphics[scale=0.5]{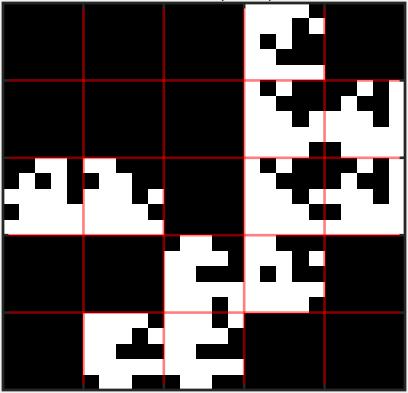} };%
\node[inner sep=0pt] at (5.5,3) { \includegraphics[scale=0.5]{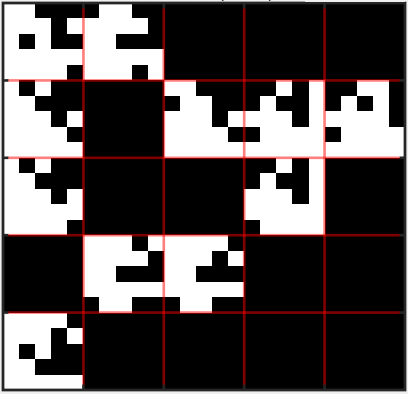} };%
\pgfmathsetmacro{\x}{-8.65}
\pgfmathsetmacro{\y}{6.1}
\foreach\i in {1,...,5} {
    \foreach\j in {1,...,5} {
        \pgfmathsetmacro{\ii}{int((6-\i)^3)}
        \pgfmathsetmacro{\jj}{int((6-\j)^3)}
        \ifthenelse{\(\i < 3 \AND \j = 3\) \OR \(\i = 2 \AND \j = 5\) \OR \(\i = 3 \AND \j > 3\) \OR \(\i = 4 \AND \j < 5\) \OR \(\i = 5 \AND \j > 1 \AND \j < 4\)}
        {\draw (\x+1.05*\i,\y-1.03*\j) node[color=white, fill=brown, fill opacity = 0.8, text opacity = 1] {1};}
        {\draw (\x+1.05*\i,\y-1.03*\j) node[color=white] {0};}
    }
}
\draw[step=1.03,black,thin,xshift=-2.6cm,yshift=0.42cm] (0,0) grid (5.15,5.15);
\pgfmathsetmacro{\x}{-3.1}
\pgfmathsetmacro{\y}{6.08}
\foreach\i in {1,...,5} {
    \foreach\j in {1,...,5} {
        \pgfmathsetmacro{\ii}{int((6-\i)^3)}
        \pgfmathsetmacro{\jj}{int((6-\j)^3)}
        \ifthenelse{\i>\j}
        {\draw (\x+1.03*\i,\y-1.03*\j) node {\ii};}
        {\draw (\x+1.03*\i,\y-1.03*\j) node {\jj};}
    }
}
\node at (-5.5,0) {$(a)$};
\node at (0,0) {$(b)$};
\node at (5.5,0) {$(c)$};
\end{tikzpicture}
\caption{OnionSearch procedure, part 1.\label{fig:onionSearchP1}. The
  binary block matrix EF (a) is permuted to maximize the dot product
  performed against the weight matrix W (b). When the local search
  reaches~$75\%$ of the optimum, the block matrix is deemed to be
  sufficiently rearranged in a top-left fashion and the procedure
  terminates (c). As evident from (b), weights are set up to push the
  maximum number of filled blocks in those layers of the onion that
  are processed first.}
\end{figure}

When~$\mat{A}$ becomes a block matrix, \emph{alternateLocalSearch}
checks the number of 1s in each nonempty block. If the blocks have a
different number of 1s, then matrix~$\mat{A}$ is not (yet) in
Kronecker form. If this happens, the local search got stuck in a (non
optimum) local minimum; to escape the minimum, the program permutes
$75\%$ of the rows/columns of~$\mat{A}$ and the
\emph{alternateLocalSearch} procedure starts again. On the other hand,
if the blocks of~$\mat{A}$ do have the same number of 1s, then the
matrix is in Kronecker form if and only if all blocks are the same. If
the blocks are different, $\mat{A}$ is processed by the
\emph{onionSearch} subroutine.

\begin{figure}[thbp]
\centering\hspace*{-1cm}\begin{tikzpicture}
\node at (-5.5,3) { \includegraphics[scale=0.5]{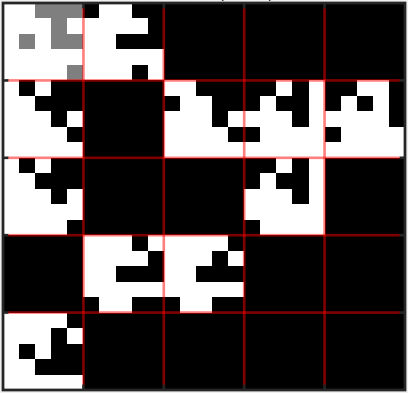} };%
\node at (0,3) { \includegraphics[scale=0.5]{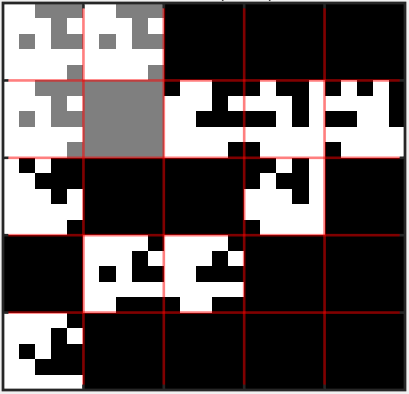} };%
\node at (5.5,3) { \includegraphics[scale=0.5]{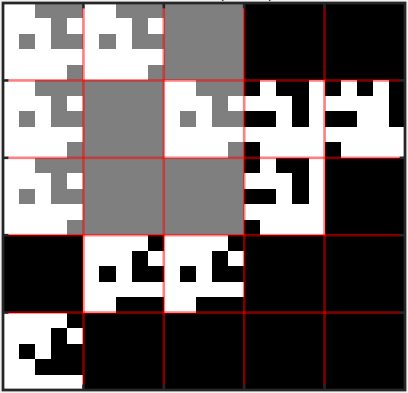} };%
\node at (-5.5,-3) { \includegraphics[scale=0.5]{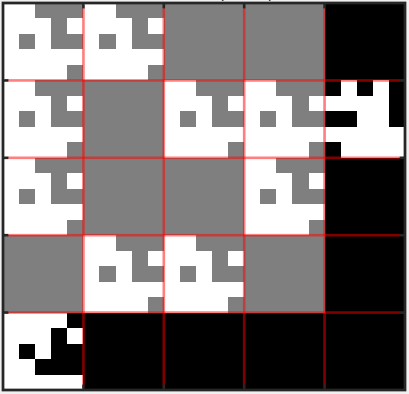} };%
\node at (0,-3) { \includegraphics[scale=0.5]{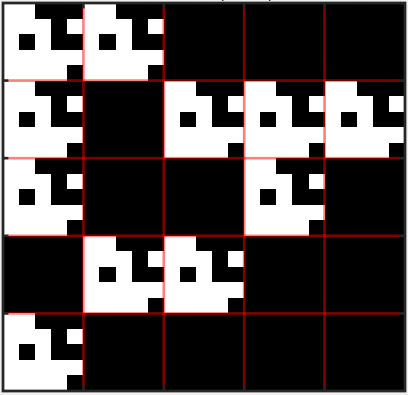} };%
\node at (3.1,-3) {$=$};
\node at (4.4,-3) { \includegraphics[scale=0.3]{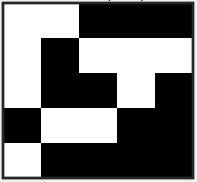} };%
\node at (5.7,-3) {$\Kron$};
\node at (7,-3) { \includegraphics[scale=0.3]{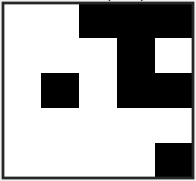} };%
\node at (-5.5,0) {$(a)$};
\node at (0,0) {$(b)$};
\node at (5.5,0) {$(c)$};
\node at (-5.5,-6) {$(d)$};
\node at (0,-6) {$(e)$};
\node at (5.5,-6) {$(f)$};
\end{tikzpicture}
\caption{OnionSearch procedure, part 2.\label{fig:onionSearchP2}. An
  example of the \emph{onionSearch} procedure. The block matrix is
  factorized layer-by-layer. The first layer, composed by a single
  block $\mat{B}_{11}$, is implicitly factorized as $\mat{B}_{11} =
  \mat{I}_1 \Kron \mat{B}_{11}$~(a). It is thus used as template
  during the local search performed on the $2 \times 2$ blocks
  submatrix, composed of four blocks~(b). As rows and columns swaps
  may only range in the second layer, the sole feasible solution is
  the one that permutes each block in accordance to the one serving as
  template. The procedure steps forward through the third~(c),
  fourth~(d) and fifth layer~(e). As the local search on the last
  layer succeeds, the problem is globally solved. The solved factors
  are shown in~(f).}
\end{figure}

The procedure \emph{onionSearch} is divided in two parts (Algorithm
\ref{alg:onionSearch}). First, the block matrix is rearranged to push
the maximum feasible number of blocks to the top-left corner, then,
starting from the top-left corner, it performs a submatrix
factorization in a layer-by-layer fashion. This second step is
designed to adopt the first nonempty block as template and to
reproduce the same layout in each other nonempty block.

Following the principles depicted in Figure~\ref{fig:onionSearchP1}, a
binary block matrix is derived from~$\mat{A}$ and it is permuted to
maximize the dot product with a weight matrix. Weights are set up to
push the maximum number of filled blocks in those layers of the onion
that are processed first.

We should now imagine the block matrix as being made of~$dimB$ layers,
starting from the top-left corner. The first layer contains one block,
the second layer contains three blocks, and so forth. The last layer
is made of blocks that belong to the last row and column of the block
matrix; refer to Figure~\ref{fig:onionSearchP1}b.  The second part of
\emph{onionSearch} tries to find a feasible factorization for
submatrices following a layer-by-layer perspective, as shown in
Figure~\ref{fig:onionSearchP2}. The local search follows the same
principles of \emph{alternateLocalSearch} but uses
metric~\metricFrob~only. Moreover, rows/columns swaps are allowed in a
limited range only. Specifically, swaps are limited to those indices
corresponding to rows and columns that fall in the unsorted layers.

It is important to note that the single block $\mat{B}_{11}$
representing the first layer is implicitly factorized as $\mat{B}_{11}
= \mat{I}_1 \Kron \mat{B}_{11}$. It is thus used as template during
the local search performed on the successive $2 \times 2$ blocks
submatrix, composed of four blocks. As rows and columns swaps may only
range in the second --- unsorted --- layer, the sole feasible solution
is the one that permutes each block in accordance to the one serving
as template. If the local search performed on the $2 \times 2$ blocks
submatrix succeeds, \emph{onionSearch} steps to the third
layer. Blocks falling in the first and in the second layer will stay
untouched as they are sorted already.  The procedure steps forward
until the last layer is processed. If the local search on the last
layer succeeds, the problem is globally solved. Otherwise,
\emph{alternateLocalSearch} proceeds testing swaps randomly to find a
feasible solution.

Finally, to handle local minima, random permutations are applied in
two more cases. First, it is possible that the list of best swaps
returned by the \emph{outsiders} procedure is exhausted without
improvements. If this condition occurs through several iterations, a
random permutation is applied to the $45\%$ of rows/columns of
$\mat{A}$. Second, if \emph{alternateLocalSearch} fails to reduce the
value of function $~\phi()$ after a maximum number of iterations
(i.e. both best swaps and default swaps are exhausted), a complete
random permutation is applied (the local search restarts from
scratch).

Note that, each time \emph{alternateLocalSearch} applies a random
permutation to escape from local minima, independently of its extent,
the procedure \emph{kronGrouping} is called again.

\section{Computational experiments}\label{sec:experiments}

\begin{figure}[ht]
\includegraphics[scale=0.5]{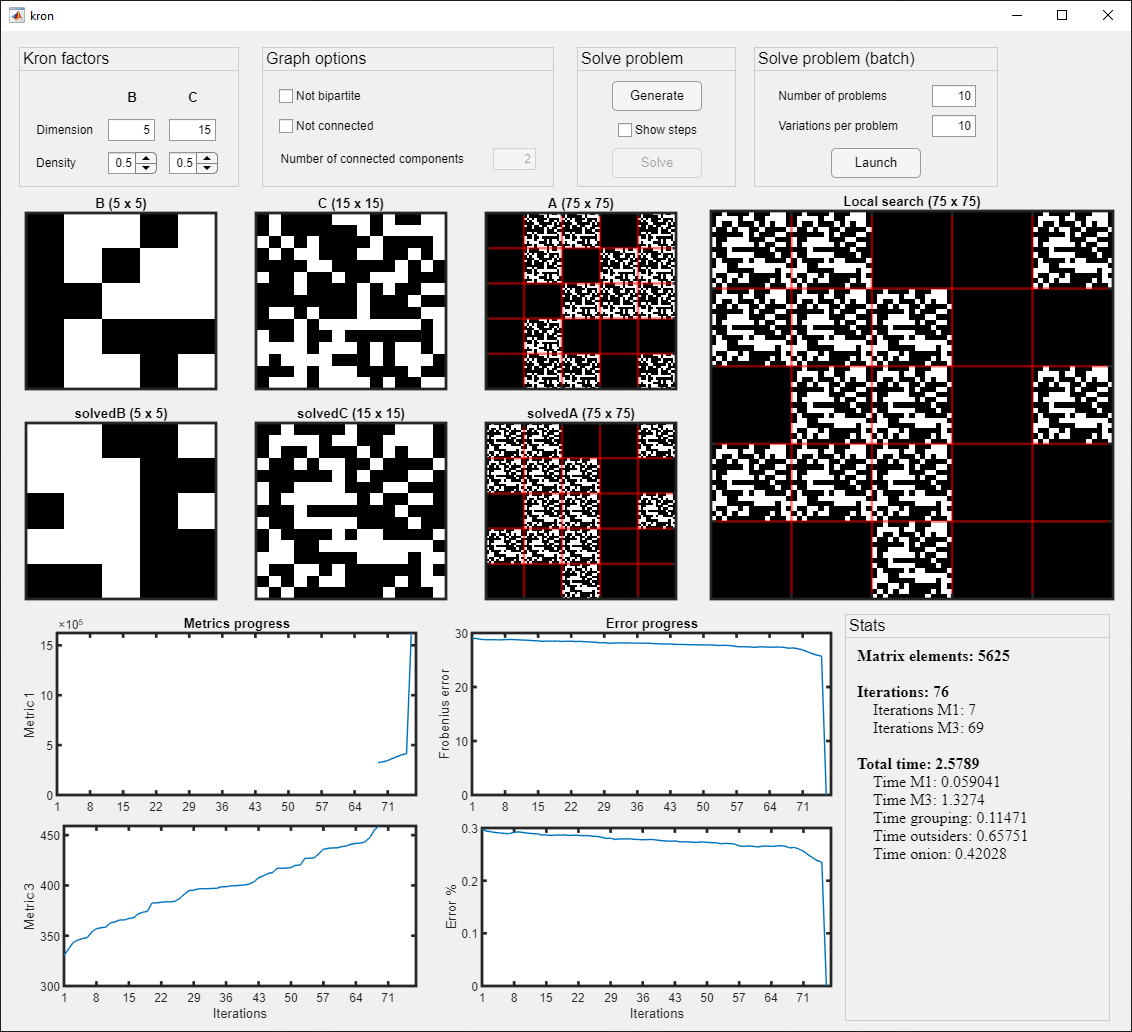}
\caption{Kronecker factorization application.\label{fig:sampleApp}.
A sample snapshot of the MATLAB application implementing the heuristic.}
\end{figure}

We implemented the heuristic described in Section~\ref{sec:algorithm}
in the MATLAB programming environment; the source code is available
online at \url{https://github.com/calderonil/kron}. The program
provides a graphical user interface (Figure~\ref{fig:sampleApp}) that
allows the user to generate a random binary matrix~$\mat{A}$ of given
size $\mathit{dimA} = \mathit{dimB} \times \mathit{dimC}$; the program
then looks for a permutation~$\mat{P}$ such that $\mat{A} =
\tr{\mat{P}}(\mat{B} \Kron \mat{C})\mat{P}$ where $\mat{B}$ and
$\mat{C}$ are binary matrices of size $\mathit{dimB} \times
\mathit{dimB}$ and $\mathit{dimC} \times \mathit{dimC}$,
respectively. $\mat{A}$ is generated starting from random $\mat{P},
\mat{B}, \mat{C}$, which are then "forgotten" and must therefore be
computed from scratch; this ensures that a factorization of $\mat{A}$
always exists. The user must provide the densities $\rho_B, \rho_C$
(fraction of 1s) of matrices $\mat{B}, \mat{C}$; the density $\rho_A$
of~$\mat{A}$ will therefore be $\rho_A = \rho_B \times
\rho_C$. Although our application allows the user to impose additional
constraints on the graph represented by the matrices (e.g.,
non-bipartitedness and/or non-connectedness), we did not impose any
constraint for the experiments described in this section.

\begin{table}[htbp]
    \centering
    \begin{tabular}{r|r|rrr|rrrrr}
        $\pmb{\rho_B, \rho_C, \rho_A}$ & \bf Size of $B$ & $\pmb{\alpha}$ & \bf Size of $C$ & \bf Size of $A$ & \bf failure \% & $\pmb{t_{min}}$ & $\pmb{t_{avg}'}$ & $\pmb{t_{avg}}$ & $\pmb{t_{max}}$ \\
        \hline
        \multirow{9}{*}{$0.5$, $0.5$, $0.25$} & \multirow{3}{*}{$5 \times 5$} & $1$ & $5 \times 5$ & $25 \times 25$ & $0.00\%$ & $0.02$ & $0.03$ & $0.06$ & $0.31$ \\
        &  & $2$ & $10 \times 10$ & $50 \times 50$ & $0.00\%$ & $0.07$ & $0.09$ & $0.13$ & $0.82$ \\
        &  & $3$ & $15 \times 15$ & $75 \times 75$ & $0.00\%$ & $0.18$ & $0.21$ & $0.27$ & $0.81$ \\
        \cline{2-10}
        & \multirow{3}{*}{$7 \times 7$} & $1$ & $7 \times 7$ & $49 \times 49$ & $0.00\%$ & $0.08$ & $0.11$ & $0.24$ & $1.28$ \\
        &  & $2$ & $14 \times 14$ & $98 \times 98$ & $0.00\%$ & $0.35$ & $0.43$ & $0.67$ & $2.55$ \\
        &  & $3$ & $21 \times 21$ & $147 \times 147$ & $1.00\%$ & $1.00$ & $1.30$ & $2.85$ & $35.72$ \\
        \cline{2-10}
        & \multirow{3}{*}{$10 \times 10$} & $1$ & $10 \times 10$ & $100 \times 100$ & $0.00\%$ & $0.52$ & $0.59$ & $1.42$ & $11.59$ \\
        &  & $2$ & $20 \times 20$ & $200 \times 200$ & $0.00\%$ & $2.50$ & $2.77$ & $3.49$ & $16.57$ \\
        &  & $3$ & $30 \times 30$ & $300 \times 300$ & $0.00\%$ & $6.71$ & $7.54$ & $8.47$ & $10.89$ \\
        \hline
        \multirow{9}{*}{$0.6$, $0.6$, $0.36$} & \multirow{3}{*}{$5 \times 5$} & $1$ & $5 \times 5$ & $25 \times 25$ & $0.00\%$ & $0.01$ & $0.03$ & $0.06$ & $0.52$ \\
        &  & $2$ & $10 \times 10$ & $50 \times 50$ & $0.00\%$ & $0.06$ & $0.08$ & $0.24$ & $3.14$ \\
        &  & $3$ & $15 \times 15$ & $75 \times 75$ & $0.00\%$ & $0.16$ & $0.19$ & $0.27$ & $1.41$ \\
        \cline{2-10}
        & \multirow{3}{*}{$7 \times 7$} & $1$ & $7 \times 7$ & $49 \times 49$ & $0.00\%$ & $0.07$ & $0.12$ & $0.45$ & $2.33$ \\
        &  & $2$ & $14 \times 14$ & $98 \times 98$ & $3.00\%$ & $0.39$ & $0.45$ & $1.20$ & $15.30$ \\
        &  & $3$ & $21 \times 21$ & $147 \times 147$ & $0.00\%$ & $0.90$ & $1.05$ & $1.80$ & $35.97$ \\
        \cline{2-10}
        & \multirow{3}{*}{$10 \times 10$} & $1$ & $10 \times 10$ & $100 \times 100$ & $0.00\%$ & $0.36$ & $0.51$ & $0.85$ & $8.42$ \\
        &  & $2$ & $20 \times 20$ & $200 \times 200$ & $0.00\%$ & $2.14$ & $2.67$ & $7.14$ & $87.05$ \\
        &  & $3$ & $30 \times 30$ & $300 \times 300$ & $0.00\%$ & $6.58$ & $7.98$ & $16.79$ & $210.95$ \\
        \hline
        \multirow{9}{*}{$0.7$, $0.7$, $0.49$} & \multirow{3}{*}{$5 \times 5$} & $1$ & $5 \times 5$ & $25 \times 25$ & $1.00\%$ & $0.01$ & $0.04$ & $0.09$ & $1.06$ \\
        &  & $2$ & $10 \times 10$ & $50 \times 50$ & $1.00\%$ & $0.07$ & $0.17$ & $0.63$ & $6.22$ \\
        &  & $3$ & $15 \times 15$ & $75 \times 75$ & $4.00\%$ & $0.14$ & $0.34$ & $0.84$ & $10.32$ \\
        \cline{2-10}
        & \multirow{3}{*}{$7 \times 7$} & $1$ & $7 \times 7$ & $49 \times 49$ & $0.00\%$ & $0.06$ & $0.18$ & $0.50$ & $3.21$ \\
        &  & $2$ & $14 \times 14$ & $98 \times 98$ & $7.00\%$ & $0.32$ & $1.18$ & $1.97$ & $18.71$ \\
        &  & $3$ & $21 \times 21$ & $147 \times 147$ & $9.00\%$ & $0.88$ & $6.21$ & $8.56$ & $102.26$ \\
        \cline{2-10}
        & \multirow{3}{*}{$10 \times 10$} & $1$ & $10 \times 10$ & $100 \times 100$ & $0.00\%$ & $0.25$ & $0.45$ & $1.84$ & $19.42$ \\
        &  & $2$ & $20 \times 20$ & $200 \times 200$ & $9.00\%$ & $2.16$ & $5.84$ & $14.91$ & $128.62$ \\
        &  & $3$ & $30 \times 30$ & $300 \times 300$ & $0.00\%$ & $5.84$ & $6.65$ & $19.91$ & $243.12$ \\
        \hline
    \end{tabular}
    \caption{Each row summarizes the performance of our heuristic on~$10$ different initial random matrices $\mat{A}$; for each matrix $\mat{A}$ we executed the heuristic~$10$ times, each one starting from a random permutation of~$\mat{A}$. $\alpha = \mathit{dimC} / \mathit{dimB}$ is the ratio between the sizes of~$\mat{C}$ and~$\mat{B}$. $t_{avg}'$ is the average of the minimum execution times of each different problem. The time spent on instances for which no solution has been found have been excluded from average, minimum and maximum execution times.}
    \label{tab:experiments}
\end{table}

The combinations of parameters used in the experiments are shown in
Table~\ref{tab:experiments}.  The density (fraction of 1s)
of~$\mat{A}, \mat{B}, \mat{C}$ are denoted as $\rho_A$, $\rho_B$ and
$\rho_C$, respectively. We consider several combinations of sizes
$\mathit{dimB}, \mathit{dimC}$ according to a parameter $\alpha$ that
denotes the ratio between the size of $\mat{C}$ and $\mat{B}$ ($\alpha
= \mathit{dimC} / \mathit{dimB}$). Intuitively, high values of
$\alpha$ denotes that the size of the factors of $\mat{A}$ are
different.

Each row of the table summarizes the result of one \emph{run}: a run
consists of~$10$ random problem instances with the chosen
parameters. Since our heuristic is sensitive to the initial
conditions, for each instance we consider~$10$ initial random
permutation of the matrix~$\mat{A}$; this is equivalent of fixing
$\mat{B}, \mat{C}$ and choosing ten different random permutation
matrices~$\mat{P}$. Therefore, each row summarizes~$10 \times 10$
executions of our program.

The program has been executed on a desktop PC with an Intel Xeon CPU
running at $3.30$ GHz with $16$~GB of RAM running Windows~10
(Matlab~R2021a). Both a \emph{single instance mode} and a \emph{batch
  mode} are provided; the former solves a single problem, while the
latter generates a set of random instances with the same parameters
($\mathit{dimA}, \mathit{dimB}, \mathit{dimC}$). To collect the
results we show in the following, the program was executed in batch
mode.

For a single instance, our program executes
\emph{alternateLocalSearch} for at most~$500$ iterations (see them
main loop of Algorithm~\ref{alg:alternateLocalSearch} in the
Appendix); when no feasible factorization is found after the maximum
allowed number of iterations, the procedure stops and the failure
count is incremented.

For each run we collect five metrics: the percentage of executions at
the end of which no factorization was found (failure \%); the minimum
and maximum execution times in seconds ($t_{min}$ and $t_{max}$,
respectively); the average execution time across all executions that
\emph{did} find the optimal solution ($t_{avg}$); the average of the
minimum execution time for each different problem ($t_{avg}'$).  More
precisely, $t_\mathit{min}'$ is computed as follows:
\begin{enumerate}
\item For each combination of parameters, we generate~$10$ random
  instances consisting of matrices $\mat{A}, \mat{B}, \mat{C}$ of the
  appropriate size and content;
\item For each instance, we generate~$10$ random permutation matrices
  $\mat{P}$; we get~$10$ variations $\mat{A} = \tr{\mat{P}}(\mat{B}
  \Kron \mat{C})\mat{P}$ that only differ by the permutation $\mat{P}$
\item We compute the minimum time to get the solution of the~$10$
  variations from the previous step;
\item $t_\mathit{ave}'$ is the average of the minimum times computed
  at the previous step.
\end{enumerate}
$t_\mathit{avg}'$ is useful because there is a significant variance
across the execution times of the~$10$ variations of the same problem
(see below). Indeed, as stated above, there might be a huge variation
in the time required to factorize a matrix~$\mat{A} = \tr{\mat{P}}
(\mat{B} \Kron \mat{C}) \mat{P}$ depending on the permutation
$\mat{P}$. Therefore, $t_{avg}'$ is the average time to solve a
problem instance if we were able to parallelize the heuristic
across~$10$ independent execution units, so that the first one that
gets the solution stops the computation.

\begin{figure}[htbp]
\centering\pgfplotsset{scaled x ticks=false}
\begin{tikzpicture}
\begin{axis}[title=Linear growth,xlabel={Number of edges in $\mat{A}$}, ylabel={Execution time (seconds)}, legend entries={$t_{avg}$,$t_{avg}'$}, legend pos=north west, xticklabel style={rotate=90,anchor=east},x label style={at={(axis description cs:0.5,-0.2)},anchor=north},]
\addplot[blue] table {img/growth1.dat};
\addplot[red] table {img/growth2.dat};
\end{axis}
\end{tikzpicture}
\caption{Dependence of the execution time on the number of 1s of
  $\mat{A}$.\label{fig:growth}. Given $\rho_A = 0.25$ and $dimB = 10$,
  this chart shows the linear growth for $t_{avg}$ and $t_{avg}'$
  depending on the size of $\mat{C}$. Showing the matrix dimension in
  the $x$-axis could be misleading, the number of edges in $\mat{A}$
  is used instead.}
\end{figure}
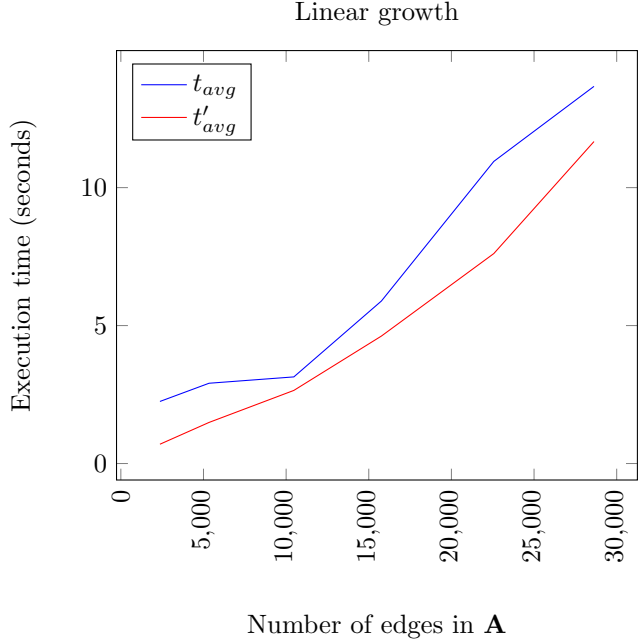

As can be observed from Table~\ref{tab:experiments}, the minimum time
to compute a solution is very low (a few seconds) for all
problems. The largest matrix $\mat{A}$ (size $300 \times 300$) can be
factored in $t_{min} = 6.71s$. It should be observed that as the
density of~$\mat{A}$ increases, the problem becomes more difficult for
our heuristic as witnessed by the increasing fraction of unsolved
instances. We also observe that there is a large variability between
the minimum and maximum times required to solve an instance; indeed we
observe that the gap between~$t_\mathit{min}$ and~$t_\mathit{max}$
becomes more than an order of magnitude, especially for large
matrices~$\mat{A}$ that are decomposed into factors of unbalanced
sizes ($\alpha = 3$). This is due to the fact that the heuristic is
sensitive to both the permutation~$\mat{P}$, and to the sequence of
swaps that are applied during the computation (the swaps are in part
generated pseudo-randomly).

We not turn our attention to the study of the dependence of the
execution time on the number of edges of the graph whose adjacency
matrix is~$\mat{A}$; the number of edges is simply the number of 1s
in~$\mat{A}$. To this aim, we performed~$60$ additional experiments
($6$ separate problem instances with~$10$ initial random permutation
each). We set $\rho_A = 0.25$ and $dimB = 10$, and we increased the
dimension of~$\mat{C}$ with a step of~$50$.  Figure~\ref{fig:growth}
shows the mean execution time $t_\mathit{avg}$ and As shown in
Figure~\ref{fig:growth}, $t_{avg}$ and $t_{avg}'$ seems to grow more
or less linearly with respect to the number of edges of~$\mat{A}$.

\section{Conclusions and future works}\label{sec:conclusions}

In this paper we presented a heuristic for decomposing a directed
graph into factors according to the direct product: given a directed,
unweighted graph~$G$ with adjacency matrix $\adj{G}$, our heuristic
searches for a pair of graphs~$G_1$ and~$G_2$ such that $G = G_1
\Direct G_2$, where $G_1 \Direct G_2$ is the direct product of~$G_1$
and~$G_2$. The heuristic proposed in this paper represents -- to the
best of our knowledge -- the first computational approach for general
directed, unweighted graphs. We provided a MATLAB implementation that
we used to run a set of computational experiments to assess the
effectiveness of our approach. Our implementation can factorize a
graph of size $300 \times 300$ in a few seconds. In a few worst-case
scenarios the time grows to a few minutes, and is due to the fact that
our heuristic is sensitive to the structure of the input; although it
may fail to find a solution, in our experiments we observed failures
in just a few instances.

We are planning to extend the heuristic along two directions: first,
to handle \emph{weighted} graphs instead of just unweighted ones;
second, to compute the \emph{approximate} Kronecker decomposition of
unweighted graphs, where (a suitable permutation of) the input matrix
$\mat{A}$ can be expressed as a Kronecker product of two smaller
matrices $\mat{B}$ and $\mat{C}$, plus an additional binary error term
$\mat{E}$, i.e., $\mat{A} = \tr{\mat{P}}(\mat{B} \Kron \mat{C})
\mat{P} + \mat{E}$.

\clearpage
\pagebreak

\begin{appendix}{Pseudocode}\label{app:pseudocode}

We provide here a more accurate description of our heuristic by means
of pseudocode.  The complete source code of the Matlab implementation
can be downloaded from~\url{https://github.com/calderonil/kron}.

\vspace{0.5cm}

\begin{algorithm}[H]
\scriptsize
\linespread{0.75}\selectfont
\SetAlgoLined
\KwIn{A, dimB, dimC}
\KwOut{solvedA, solvedB, solvedC}
 
 bestA = A\;
 metric = \textsc{var}\;
 A = kronGrouping(A)\;
 \tcc{Main loop}
 \While{iter $\leq$ maxiter}
 { 
    iter++\;
    \If{$\sim$blockMatrix}{bestSwaps = outsiders(bestA)\;}
  
    \If{$\sim$cornerizedMatrix $\wedge$ blockMatrix}
    {
        \If{$\sim$balancedBlocks}
        {
            \tcc{Blocks have a different number of 1s, unfeasible factorization}
            bestA = randPerm(bestA,75\%)\;
            bestA = kronGrouping(bestA)\;
            blockMatrix = \textbf{false}\;
        }
    }
  
    \If{$\sim$cornerizedMatrix $\wedge$ blockMatrix}
    {
        [bestA, success] = onionSearch(bestA)\;
        \eIf{success}{\textbf{break}\;}{\textbf{continue}\;}
    }
  
    swaps = [bestSwaps,baseSwaps]\tcp*{baseSwaps: all comb. of \{1\dots dimA\} two at a time}
  
    \For{i=1 \KwTo len(swaps)}
    {
        \If{bestSwaps exhausted}
        {
            baseSwaps = randPerm(baseSwaps,100\%)\;
            \If{metric == \textsc{frob}} 
            {
                bestA = randPerm(bestA,45\%)\;
                bestA = kronGrouping(bestA)\;
                blockMatrix = \textbf{false}\;
                cornerizedMatrix = \textbf{false}\;
            }
            \If{metric == \textsc{var}}{metric = \textsc{frob}\;}
        }
  
        curSwap = swaps(i)\;
        testA = swap(bestA, curSwap)\;
  
        \If{improved}
        {
            bestA = testA\;
            \If{metric == \textsc{var} $\wedge$ curVal $\geq$ threshold}{metric = \textsc{frob}}
            \textbf{break};
        }
    }
  
    [solvedB, solvedC] = nearestKronProduct(bestA)\;
  
    \If{$\sim$improved}
    {
        nRestarts++\;
        \eIf{nRestarts $\bmod$ perturbateEvery == 0 } 
        {
            bestA = randPerm(bestA,100\%)\;
            bestA = kronGrouping(bestA)\;
            blockMatrix = \textbf{false}\;
            cornerizedMatrix = \textbf{false}\;
            metric = \textsc{var}\;
        }
        {
            swapMetric()\;
        }
    }
  
    \If{nRestarts $>$ maxRestarts $*$ perturbateEvery $\vee$ error $< 0.000001$}{\textbf{break\;}}
  
 }
 
 \KwRet bestA, solvedB, solvedC\;
 
\caption{Alternate local search}\label{alg:alternateLocalSearch}
\end{algorithm}

\clearpage
\pagebreak

\begin{algorithm}[H]
\small
\linespread{0.75}\selectfont
\SetAlgoLined
\KwIn{A}
\KwOut{groupedA}
\For{i=1 \KwTo dimA}
{
    \For{j=1 \KwTo dimA}
    {
        \eIf{i==j}{MR(i,j)=MC(i,j)=0\;}
        {
            \tcc{Simil computes the similarity between two rows/cols relying on the dot product. To weight zero-based and one-based similarity, both A and 1-A are considered.}
            MR(i,j) = simil(A(i,:),A(j,:)) + simil(1-A(i,:),1-A(j,:))\;
            MC(i,j) = simil(A(:,i),A(:,j)) + simil(1-A(:,i),1-A(:,j))\;
        }
    }
}

\tcc{Six different permutations are tested, each one relying on a linear combination of rows similarity and columns similarity.}
\For{w=0 \KwTo 1 \emph{\bf step} 0.2}
{
    MS = w * MR + (1-w) * MC\; 
    \For{i=1 \KwTo dimA}
    {
        pivot = first unused index\;
        scan MS(pivot,:) for the (dimC-1) best indexes and add them to the current permutation\;
    }
}

\KwRet best permutation according to metric = \textsc{frob}

\caption{Kron grouping}\label{alg:kronGrouping}
\end{algorithm}

\vspace{0.5cm}

\paragraph{Comment to Algorithm~\ref{alg:outsiders}}
At the beginning, \emph{outsiders} identifies which block of~$\mat{A}$
should receive 1s (because it already has more 1s than average) and
which block should lose 1s. Then, two matrices are produced,
$\mat{WR}$ and~$\mat{WC}$. $\mat{WR}$ contains the number of 1s found
in each portion of a row of length~$dimC$ --- the portion of the row
corresponding to a given block. If the block should receive 1s, the
number of zeros is counted instead. The second matrix~$\mat{WC}$
contains the same information computed column-wise.  A third
matrix~$\mat{MS}$ is computed in order to find the best swap: the
procedure loops on each row/column on a block-by-block basis and
multiplies the elements of~$\mat{WR}$ and~$\mat{WC}$ corresponding to
blocks that are in opposite conditions, i.e. one receives and the
other loses 1s. The larger the values of~$\mat{MW}$, the better. This
product is tuned according to the density of the block that receives
1s. However, moving elements in blocks that already have a high number
of 1s would produce blocks with too many 1s, which are unlikely to
appear in a feasible factorization.  Elements of~$\mat{MS}$ are
scanned and ordered to form the list of more promising swaps.
If~$\mat{MS}$ has zero elements only, there are no 1s on the blocks
that should lose 1s. Therefore, according to
Definition~\ref{def:blockMatrix}, the matrix is a block matrix.

\clearpage
\pagebreak

\begin{algorithm}[H]
\small
\linespread{0.75}\selectfont
\SetAlgoLined
\KwIn{A}
\KwOut{bestSwaps}

BL = blockMatrix(A)\tcp*{derive the dimB x dimB block matrix} 
S = blocksSum(BL)\tcp*{count the total number of 1s in each block}
D = blocksDensity(BL)\tcp*{computes the density of 1s in each block}

\ForEach{B \emph{\bf in} BL}
{
    \For{k=1 \KwTo dimC}
    {
        sumRow = sum(B(k,:))\;
        sumCol = sum(B(:,k))\;
        \eIf{S(B) $<$ avg(S)} 
        {
            \tcc{The block is likely to be emptied, count the number of 1s}
            WR(((row(B)-1)*dimC)+k,col(B)) = sumRow\;
            WC(row(B),((col(B)-1)*app.dimC)+k) = sumCol\;
        }
        {
            \tcc{The block is likely to be filled, count the number of zeros}
            WR(((row(B)-1)*dimC)+k,col(B)) = -(dimC-sumRow)\;
            WC(row(B),((col(B)-1)*app.dimC)+k) = -(dimC-sumCol)\;
        }
    }
}

\For{i=1 \KwTo dimA} 
{
    p = blockIdx(i)\;
    \For{j=1 \KwTo dimA}
    {
        q = blockIdx(j)\;
        \If{i and j fall in different blocks}
        {
            \For{b=1 \KwTo dimB}
            {
                \If{BL(p,b) is a block to be emptied and BL(q,b) a block to be filled}
                {
                    MS(i,j) = MS(i,j) + (WR(i,b) * WR(j,b) * (1-D(q,b))$^2$)\;
                }
                \If{BL(p,b) is a block to be filled and BL(q,b) a block to be emptied}
                {
                    MS(i,j) = MS(i,j) + (WR(i,b) * WR(j,b) * (1-D(p,b))$^2$)\;
                }
                \If{BL(b,p) is a block to be emptied and BL(b,q) a block to be filled}
                {
                    MS(i,j) = MS(i,j) + (WC(b,i) * WC(b,j) * (1-D(b,q))$^2$)\;
                }
                \If{BL(b,p) is a block to be filled and BL(b,q) a block to be emptied}
                {
                    MS(i,j) = MS(i,j) + (WC(b,i) * WC(b,j) * (1-D(b,p))$^2$)\;
                }
            }
        }
    }
}

\eIf{sum(MS)==0}
{
    blockMatrix = \textbf{true}\;
    \KwRet \textbf{null}\;
}
{
    \KwRet a list of swaps (i,j) in ascending order of MS(i,j)\;
}

\caption{Outsiders}\label{alg:outsiders}
\end{algorithm}

\clearpage
\pagebreak

\begin{algorithm}[H]
\small
\linespread{0.75}\selectfont
\SetAlgoLined
\KwIn{A}
\KwOut{A}

\tcc{------- PART1: rearrange the filled blocks in a top-left fashion ---------}
BL = blockMatrix(A)\;
EF = checkFilledBlocks(BL)\tcp*{derive a boolean block matrix; 0: empty block, 1: o.w.}
W = weightMatrix(dimB)\tcp*{the max weight corresponds to the top-left block} 

cornerizeVal = sum(dotProduct(EF,W))\; 

\For{iter = 1 \KwTo maxiterCornerize}
{
    \For{i=1 \KwTo len(swaps)}
    {
        curSwap = swaps(i)\;
        testEF = swap(EF, curSwap)\;
        testVal = sum(dotProduct(testEF,W))\;
        
        \If{testVal $>$ cornerizeVal}
        {
            cornerizeVal = testVal\;
            EF = testEF\;
            BL = swap(BL, curSwap)\;
            \textbf{break};
        }
    }
    
    \If{cornerizeVal $>$ 0.75 * cornerizeOptVal}
    {
        \textbf{break};
    }
    \If{$\sim$improved}
    {
        BL = randPerm(BL,55\%)\;
        Apply the same permutation to EF\;
    }
}
cornerizedMatrix = \textbf{true}\;

\tcc{----- PART2: perform a layer-by-layer Kron factorization on submatrices --}
A = fromBlocks(BL)\;
settledLayers = 1\; 
\For{curlayer = 2 \KwTo dimB}
{
    \tcc{In localSearchSubmatrix swaps are limited to the range of blocks [settledLayers+1, curLayer]. Moreover, only the Frobenius metric is used.}
    [success, A] = localSearchSubmatrix(A, curLayer, settledLayers)\;
    \If{success}
    {
        settledLayers = curLayer\;
    }
}

\KwRet success, A\;

\caption{Onion search}\label{alg:onionSearch}
\end{algorithm}

\vspace{0.5cm}
\paragraph{Comment to Algorithm \ref{alg:onionSearch}}
The procedure \emph{onionSearch} is divided in two parts. First, the
block matrix is rearranged to push the maximum feasible number of
blocks to the top-left corner to facilitate the second step. A binary
block matrix~$\mat{EF}$ is derived from~$\mat{A}$ and it is permuted
to maximize the dot product with the weight matrix~$\mat{W}$. To find
a convenient permutation, random swaps are applied following the same
principle of the main local search. When the procedure reaches the
$75\%$ of the optimum, the \emph{block matrix} is deemed to be
sufficiently rearranged in a top-left fashion and the procedure
terminates.  The second part of \emph{onionSearch} tries to find a
feasible factorization for submatrices following a layer-by-layer
perspective. The local search follows the same principles of
\emph{alternateLocalSearch} but uses metric~\metricFrob~only through
the subroutine \emph{localSearchSubmatrix}.  It is important to note
that the single block $\mat{B}_{11}$ representing the first layer is
implicitly factorized as $\mat{B}_{11} = \mat{I}_1 \Kron
\mat{B}_{11}$. It is thus used as template during the local search
performed on the successive $2 \times 2$ blocks submatrix, composed of
four blocks. The procedure steps forward until the last layer is
processed. If the local search on the last layer succeeds, the problem
is globally solved. Otherwise, \emph{alternateLocalSearch} proceeds
testing swaps randomly to find a feasible solution.

\end{appendix}

\clearpage
\pagebreak

\bibliographystyle{plainnat}
\bibliography{kron}

\end{document}